\documentclass[11pt]{article}
\usepackage{amsmath}
\usepackage{amssymb}
\usepackage{mathptmx}
\usepackage{amsfonts}
\usepackage[mathscr]{eucal}
\usepackage[dvips]{graphicx}
\usepackage{color}

\newcommand{\sym}{\, \textrm{sym}}

\newcommand{\bos}{\textrm{b}}

\newcommand{\fin}{\textrm{fin}}

\newcommand{\Rd}{\mathbb{R}^{\, d} }
\newcommand{\Rdn}{\mathbb{R}^{ \, d n} }

\newcommand{\Rdk}{\mathbb{R}^{\, d}}
\newcommand{\Rdx}{\mathbb{R}^{\, d}}

\newcommand{\Rdnk}{\mathbb{R}^{\, dn}}

\newcommand{\dk}{d \mathbf{k} }

\newcommand{\Fb}{\mathscr{F}_{\, \textrm{b}}  }

\newcommand{\Psin}{\Psi^{\,(n)}}  
\newcommand{\Phin}{\Phi^{\,(n)}}  
  
\newcommand{\HI}{H_{\textrm{I}}}

\newcommand{\mbf}[1]{\ensuremath{\mathbf{#1}}}
\newcommand{\mbb}[1]{\ensuremath{\mathbb{#1}}}
\newcommand{\ms}[1]{\ensuremath{\mathscr{#1}}}

\newcommand{\sqzb}[1]{\ensuremath{d\Gamma_{\textrm{b}}({#1}) }}

\newcommand{\fI}{f_{{}_{\textrm{I}}}}

\newcommand{\cI}{c_{{}_{\textrm{I}}}}
\newcommand{\dI}{d_{\,{}_{\textrm{I}}}}

\newcommand{\Bd}{B_{d}}

\newcommand{\phiS}{\phi_{{}_{\textrm{S}}}}
\newcommand{\piS}{\pi_{{}_{\textrm{S}}}}

\newcommand{\FPhiPsi}{F_{{}_{\Phi, \Psi}}}

\newcommand{\FUPhiUPsi}{F_{{}_{\Phi_{{}_U}, \Psi_{{}_U}}}}
\newcommand{\FOmegaUPhiUPsi}{F_{{}_{\omega, \, \Phi_{{}_U}, \Psi_{{}_U}}}}

\newcommand{\pPhiPsi}{\phi_{{}_{\, \textrm{cl}, \, \Phi, \Psi}}}

\newcommand{\pUPhiUPsi}{\phi_{{}_{\, \textrm{cl}, \,  \Phi_{{}_U}, \Psi_{{}_U}}}}
\newcommand{\pUPsiUPsi}{\phi_{{}_{\, \textrm{cl}, \,  \Psi_{{}_U}, \Psi_{{}_U}}}}
\newcommand{\FzeroPhiPsi}{F_{{}_{\, 0, \,\Phi, \Psi}}}
\newcommand{\FzeroPsiPhi}{F_{{}_{\, 0, \,\Psi, \Phi}}}
\newcommand{\FPsiPhi}{F_{{}_{\Psi, \Phi}}}

\newtheorem{theorem}{Theorem}[section]
\newtheorem{proposition}[theorem]{Proposition}
\newtheorem{lemma}[theorem]{Lemma}
\newtheorem{corollary}[theorem]{Corollary}
\newtheorem{remark}{Remark}[section]

\begin{document}
\begin{center}
{\LARGE Construction of  solutions of the classical field equation for  a  massless Klein-Gordon field  coupled to a  static source}
 \\
 $\;$ \\
 {\large Toshimitsu Takaesu }  \\
 $\;$ \\
\textit{Faculty of Science and Technology, Gunma University,\\ Gunma, 371-8510, Japan }
\end{center}

\begin{quote}
\textbf{Abstract} In this paper, we consider a system of a  massless  Klein-Gordon field coupled to a static source. The  total Hamiltonian is a self-adjoint operator on a boson Fock space. We consider   annihilation  operators  in  the Heisenberg picture and define a sesquilinear form. Under infrared regularity conditions, it is proven  that  the sesquilinear  form is  a solution of the classical field equation.  \\
\end{quote}
{\small
key words : Quantum field theory, Fock space, Self-adjoint operator}.

\section{Introduction}
In this paper, we consider a system of a massless Klein-Gordon  field coupled to a static source.    Let $\phi_{\textrm{cl}}= \phi_{\textrm{cl}} (t,\mbf{x})$, $(t $, $\mbf{x}) \in  \mathbb{R} \times \Rd$, be the classical field function. The Lagrangian density for the classical field is given by  
 \[
\ms{L}_{\textrm{cl}}= \frac{1}{2}(\partial_t \phi_{\textrm{cl}})^2 -\frac{1}{2}| \nabla \phi_{\textrm{cl}} |^2
 + \rho (\mbf{x}) \phi_{\textrm{cl}} ,
\]
where $\nabla=(\partial_{x_j})_{j=1}^d$ and $\rho  = \rho(\mbf{x}) $ is the density function of the static source. The Euler-Lagrange equation yields that
\begin{equation}
(\partial_t^2 - \triangle ) \phi_{\textrm{cl}}  (t,\mbf{x}) 
= \rho (\mbf{x} ),  \notag 
\end{equation}
where $\triangle=\sum\limits_{j=1}^d \partial_{x_j}^2$. The main purpose in this paper is to construct  solutions of the above classical field equation   from  the quantized  field.  The  Hilbert space for the system  is defined  by a boson Fock space. The quantized total Hamiltonian  is given by 
\begin{equation}
H \, = \,    \int_{\Rd} \omega (\mbf{k} ) a^{\dagger} (\mbf{k}) a (\mbf{k}) \dk  - \int _{\Rd }\frac{1}{\sqrt{2 \omega(\mbf{k})}}
\left( \frac{}{} \hat{\rho} (-\mbf{k})  a(\mbf{k}) + \hat{\rho} (\mbf{k})  a^{\dagger} (\mbf{k}) \right)  \dk  . \label{total}
\end{equation}
 Here $ \omega (\mbf{k}) =|\mbf{k}|$ denotes the dispersion relation,  $a^{\dagger} (\mbf{k})$ the   creation operator, $ a (\mbf{k}) $ the annihilation operator  and  $\hat{\rho} (\mbf{k}) $ the Fourier transform of $\rho (\mbf{x})$. Under  momentum cutoff conditions for $\hat{\rho} (\mbf{k}) $, $H$ is a self-adjoin operator on  the  boson Fock space. The type of the   above Hamiltonian  is  called  van Hove Hamiltonian. 

Let us define  a  sesquilinear form  by 
\begin{equation}
\pPhiPsi (t, \mbf{x})
=  \frac{1}{\sqrt{2 \pi}^{\,d}} \int_{\Rd} 
\frac{1}{\sqrt{2 \omega (\mbf{k})}} 
\left(  \FPhiPsi (t, \mbf{k}) e^{\,i\, \mbf{k} \cdot \mbf{x}} 
+ \FPsiPhi (t, \mbf{k})^{\ast} e^{\,-i\, \mbf{k} \cdot \mbf{x}}
 \right) \dk  , 
\end{equation}
where $\FPhiPsi (t, \mbf{k})=(\Phi, e^{\,itH}  a( \mbf{k} ) e^{-itH } \Psi )$. Note that we consider the annihilation  operator $e^{\,itH}  a( \mbf{k} ) e^{-itH }$ in  the Heisenberg picture.
We  additionally assume momentum cutoff conditions for  $\hat{\rho} (\mbf{k}) $, which include  infrared regularity conditions, and  introduce a unitary operator $U$. Then the unitary transformation of $H$ is   a sum of the free Hamiltonian and a constant number. In this sense, van Hove model is called exactly solvable model. Let  $d \geq 3$.  Using the unitary transformation and vectors of the form $\Theta_{\, U} = U \Theta  $,    it is proven  that 
\begin{equation}
  \left(   \partial_t^2 - \triangle \right) \pUPhiUPsi (t, \mbf{x} ) 
\, = \, \rho (\mbf{x}) .
\end{equation}

 We may think that 
the result in this paper shows a relation between  a classical field and the quantized field. It also  can be regarded as a justifiable feature of field quantizations.
Quantum field theory has  divergent problems in itself, however,  
 mathematically rigorous results have been obtained by many researchers. 
  For the recent  research on van Hove models, the classification of  ultraviolet and infrared divergent properties and scattering theory were investigated in \cite{De03}. Various  mathematical features of interacting  quantum fields, which include  van Hove models,  were   considered in  \cite{De14}.

This paper is organized as follows. In Section 2, basic properties of boson Fock spaces and their operators are reviewed, the definition of the total Hamiltonian is given, and  the main theorem is stated. In section 3,  the proof of  the main theorem is given. 

%%%%%%%%%%%%%%%%%%%%%%%%%%%%%%%%%%%%%%%%%%%%%%%%%%%%%%%%%%%%%
%%%%%%%%%%%%%%%%%%%%%%%%%%%%%%%%%%%%%%%%%%%%%%%%%%%%%%%%%%%%%%
%%%%%%%%%%%%%%%%%%%%%%%%%%%%%%%%%%%%%%%%%%%%%%%%%%%%%%%%%%%%%%%
\section{Definitions and Main Result}
\subsection{Preliminaries}
Let $\ms{H}$ be a complex Hilbert space. The  inner product   is denoted by  $(\Phi, \Psi )$  which is   linear in the second argument and the norm  by $\| \Psi \| $. For  a linear operator $X$ on $\ms{H}$, the domain of $X$ is denoted by $D(X)$, the  adjoint    by $X^{\, \ast}$, and its closure  by $\overline{X}$ if   $X$ is  a closable. \\
\subsection{Fock Spaces} 
In this subsection, we review  basic properties of Fock spaces. (refer to  e.g.,  \cite{Arai}, \cite{RSI},\cite{RSII}) \\
$\;$ \\
\textbf{[i] Boson Fock Space } \\
Let $ d  \in \mbb{N} $ be a spatial dimension.  
The Hilbert space for symmetric $n$ particles in $\Rd$ is given by $L^2_{\sym} (\Rdn)  $ which consists of  all vectors  $ \Psin \in L^2 (\Rdn ) $ such that  for all $ \sigma \in \mathfrak{S}_n$, 
$\Psin (\mbf{k}_1 , \cdots , \mbf{k}_n ) = 
\Psin (\mbf{k}_{\sigma (1)} , \cdots , \mbf{k}_{\sigma (n)} )  , \; \textrm{a.e.}
\;  (\mbf{k}_1 , \cdots , \mbf{k}_n ) \in \Rdnk $, where  $ \mathfrak{S}_n $ denotes the  symmetric group of  degree $n$.  The  boson Fock space over $L^2 (\Rd )  $  is defined by 
\[
\Fb = \bigoplus_{n=0}^{\infty} L^2_{\sym} (\Rdnk) 
\]
with   $ L^2_{\sym} (\mathbb{R}^0 ) = \mbb{C} $.
Let $\Phi = \{ \Phin \}_{n=0}^{\infty}$, $\Psi =\{ \Psin \}_{n=0}^{\infty} \in \Fb$. The inner product of $\Fb$ is given by
\[
(\Phi, \Psi ) =  \sum_{n=0}^{\infty} ( \Phin , \Psin ) .
\]
The finite particle subspace $\ms{F}_{\bos, \fin}$  is the set of
all  vectors $\Psi  \in \Fb $   which satisfy that there exists $N \in \mbb{N}$ such that for all $n  > N$, $ \Psi^{\, (n)} =0 $. Unless otherwise specified,  the domain of a linear operator $X$ on $\Fb$ is defined by
\[
\ms{D}(X) = \left\{ \Psi  \in \Fb \left| \frac{}{} \right. 
 \sum_{n=0}^{\infty} \| (X \Psi )^{(n)} \|^2  \, < \, \infty \right\}.
\]

$\;$ \\
\textbf{[ii] Annihilation and Creation Operators }  \\
The annihilation operator $a(f)$  smeared with $f \in L^{2}(\Rdk )$ is defined by
\[
(a(f) \Psi )^{\, (n)} (\mbf{k}_1 , \cdots , \mbf{k}_n)
= \sqrt{n+1} \int_{\Rd} f(k)^{\ast} \,  \Psi ^{\, (n+1)} (\mbf{k} ,\mbf{k}_1 , \cdots \mbf{k}_n   ) \dk , \quad n \geq 0 ,
\]
where  $z^{\ast}$ denotes the complex conjugate of  $ z \in \mbb{C}$. 
The creation operator $a^{\dagger}(g)$  smeared with $g \in L^{2}(\Rdk )$ is defined by
\[
(a^{\dagger}(g) \Psi )^{\, (n)} (\mbf{k}_1 , \cdots , \mbf{k}_n) = 
 \frac{1}{ \sqrt{n}} 
  \sum\limits_{j=1}^{n} g(\mbf{k}_j ) \Psi^{\, (n-1)}(\mbf{k}_1 , \cdots ,\mbf{k}_{j-1} , 
\mbf{k}_{j+1}  , \cdots , \mbf{k}_n) ,  \quad n \geq 1  ,
 \]
and $(a^{\dagger}(g) \Psi )^{\, (0)}=0$.
It holds that   
 $ a(f) = (a^{\dagger } (f))^{\ast}$.  The creation and annihilation operators satisfy the canonical commutation relations  
\begin{align}
  &{[} a(f) , a^{\dagger } (g) {]} = (f,g)   , \label{CCR1} \\
   &{[} a(f) , a (g) {]} =  {[} a^{\dagger}(f) , a^{\dagger } (g) {]} = 0 , \label{CCR2} 
\end{align}
 on $\ms{F}_{\bos , \fin }$.  The Segal field operators  and their conjugate operators are defined by 
\begin{align*}
&\quad \phiS (f)= \frac{1}{ \sqrt{2}} \left(  a(f) + a^{\dagger} (f) \right) , \qquad \; \; f \in 
L^2 ( \Rdk ),  \\
&\quad \piS (g)= \frac{i}{ \sqrt{2}} \left(  - a(g) + a^{\dagger} (g) \right) , \qquad g \in
 L^2 ( \Rdk ) .
\end{align*}
From (\ref{CCR1}) and (\ref{CCR2}), it follows that  
\begin{align}
&  [ \phiS (f) , \, \piS (g) ] =  i \, \textrm{Re} (f,g) , \label{CCR3} \\
&   [ \phiS (f) , \,  \phiS (g) ] =  {[} \piS (f) , \piS (g) {]} = 0 ,  \label{CCR4}
\end{align}
 on $\ms{F}_{\bos , \fin }$. It holds that $\phiS (f)$ and  $\piS (g)$ are essentially self-adjoint. From (\ref{CCR1}) and (\ref{CCR2}), it follows that 
\begin{equation}
 e^{\,i \,\overline{\piS (g) }} a(f) e^{\,-i \,\overline{\piS (g)} } \,=\, a(f) + \frac{1}{\sqrt{2}} (f, g). \label{WCR1} 
 \end{equation}
$\;$ \\
\textbf{[iii] Second Quantization} \\
Let $T = T(\mathbf{k})$ be a non-negative and Borel measurable function. The second quantization $\sqzb{T}$ of $T$ is defined by  
\[
(\sqzb{T} \Psi )^{\, (n)} (\mbf{k}_1 , \cdots , \mbf{k}_n) =  
 \sum_{j=1}^{n} T (\mbf{k}_j ) \Psin (\mbf{k}_1 , \cdots , \mbf{k}_n) , \quad  n \geq 1 ,
\] 
and $(\sqzb{T} \Psi )^{\, (n)} =0$. 
Let $f \in L^{2} (\Rdk )$ with $T^{-1/2} f \in  L^{2} (\Rdk )$. It holds that for all $\Psi \in \ms{D} (\sqzb{T}^{1/2})$,
\begin{align}
&\| a(f) \Psi \| \leq \|T^{\,-1/2} f \| \, \| \sqzb{T}^{1/2} \Psi  \| ,  \label{8/19.a} \\
& \| a^{\dagger}(f) \Psi \| \leq \|T^{\,-1/2} f \| \, \| \sqzb{T}^{1/2} \Psi  \| + \| f\| \, \| \Psi \|  \label{8/19.b}. 
\end{align}
For all $f  \in L^{2} (\Rdk )$ satisfying $Tf   \in L^{2} (\Rdk )$, it holds that
\begin{align}
&[  \sqzb{T} , a (f) ] = - a(Tf ) , \label{8/19.c} \\
&[  \sqzb{T} , a^{\dagger} (f) ]  = a^{\dagger} (Tf) ,  \label{8/19.d}
\end{align}
 on $\ms{F}_{\bos , \fin}$. 
By (\ref{8/19.c}) and (\ref{8/19.d}), 
it  follows that 
\begin{equation}
 e^{\, it \sqzb{T} } a(f) e^{\, -it \sqzb{T} } = a (e^{\, itT} f)  \label{8/23.a} 
\end{equation}
and 
\begin{equation}
e^{\,i \,\overline{\piS (g) }} \sqzb{T} e^{\,-i \,\overline{\piS (g)} } \,=\, 
\sqzb{T}+ \phiS (Tg) + \frac{1}{2} (g, Tg ) .  \label{diag1}
\end{equation}
\subsection{Hamiltonian and Main Theorem }
The total  Hamiltonian $H$ on $\Fb$ is given by 
\begin{equation}
H = H_0 + \HI ,
\end{equation}
where $H_0= \sqzb{\omega}$ with $\omega(\mbf{k}) = | \mbf{k} |$ and  $\HI = - \phiS (\fI )$ with $\fI (\mbf{k}) =  \frac{\hat{ \rho }(\mbf{k})}{\sqrt{ \omega(\mbf{k})}}$.  

Suppose the condition below.
\begin{quote}
\textbf{(A.1)} $\rho \in L^1 (\Rdx )$  and 
$\frac{\hat{\rho}}{ \sqrt{\omega}^{\,l}} \in L^2 (\Rdk ) $, $l=1,2$.
\end{quote}
$\;$ \\
We quickly check the self-adjointness of $H$. By (\ref{8/19.a}) and (\ref{8/19.b}), we have
\begin{equation}
\|\phiS (\fI ) \Psi  \| 
\leq \sqrt{2} \left\|  \frac{\fI}{\sqrt{\omega}} \right\| \, 
\|  H_0^{1/2} \Psi \| +  \frac{1}{\sqrt{2}} \| \fI \| \,  \| \Psi \|. \notag
\end{equation}
From  spectral decomposition theorem, we obtain 
$\|  H_0^{1/2} \Psi \| \leq \epsilon \| H_0 \Psi \| + \frac{1}{2 \epsilon} \| \Psi \| $, $\epsilon > 0$. Therefore, 
\[
\| \HI  \Psi  \| 
\leq  \cI \epsilon  \|  H_0 \Psi \| +  \dI (\epsilon) \| \Psi \| ,
\]
where $ \cI= \sqrt{2} \left\|  \frac{\fI}{\sqrt{\omega}} \right\| $ and
$ \dI (\epsilon )= \frac{1}{ \sqrt{2}\, \epsilon} \left\|  \frac{\fI}{\sqrt{\omega}} \right\|
+ \frac{1}{\sqrt{2}} \| \fI\|$. Taking   $ \epsilon > 0 $ such that $  \epsilon < \frac{1}{\cI}$, the   Kato-Rellich theorem yields that $H$ is self-adjoint on $\ms{D}(H_0)$.

  Let 
\begin{equation}
\FPhiPsi (t, \mbf{k}) = (\Phi, e^{\,itH}  a( \mbf{k} ) e^{-itH } \Psi ) , \notag
\end{equation}
where $a(\mbf{k})$ is the operator  kernel of the annihilation operator, which is defined in Section \ref{OperatorKernel}.  
Let
\begin{equation}
\pPhiPsi (t, \mbf{x})
=  \frac{1}{\sqrt{2 \pi}^{\,d}} \int_{\Rd} 
\frac{1}{\sqrt{2 \omega (\mbf{k})}} 
\left(  \FPhiPsi (t, \mbf{k}) e^{\,i\, \mbf{k} \cdot \mbf{x}} 
+ \FPsiPhi (t, \mbf{k})^{\ast} e^{\,-i\, \mbf{k} \cdot \mbf{x}}
 \right) \dk  . \notag
\end{equation}

$\;$ \\
Suppose the  condition below. 

\begin{quote}
\textbf{(A.2)}  
$\frac{\hat{\rho}}{ \sqrt{\omega}^{\,3}}\in  L^2 (\Rdk ) $, $\hat{\rho} \in L^1 (\Rdk )$ and  $
  \frac{\hat{\rho}}{ \omega^{\,2}} \in L^1 (\Rdk )$. 
\end{quote}

\begin{remark} \normalfont
 If $d=3$,  it holds that  $ \int_{\mbb{R}^3} \frac{1}{\omega  (\mbf{k})^3} \dk = \infty $. Hence   the condition   
 $\frac{\hat{\rho}}{ \sqrt{\omega}^{\,3}}
 \in     L^2 (\Rdk )$ is called  infrared regularity condition.
 \end{remark}
$\;$ \\
By \textbf{(A.2)}, we can define a unitary operator   
\[
U = e^{\, -i \,\overline{\piS \left( \frac{\fI}{\omega} \right)}} .
\]
From (\ref{diag1}),  it holds that 
\begin{equation}
U^{\ast} H U = H_{0} - \frac{1}{2} (\frac{\fI}{\omega} , \fI ) . \label{diagH}
\end{equation}
Here we state the main theorem. 
\begin{theorem} \label{MainTheorem} \normalfont 
Assume \textbf{(A.1)}, \textbf{(A.2)} and $d \geq 3$. Then for all
 $\Phi , \Psi \in \ms{D} (H_{0}^{\frac{d}{2}+2})$ with $(\Phi , \Psi ) = 1$,
\begin{equation}
\left(   \partial_t^2 - \triangle \right) \pUPhiUPsi (t, \mbf{x} ) 
\, = \, \rho (\mbf{x}) , \label{main}
\end{equation}
where $\Phi_{\, U} = U \Phi $ and $\Psi_{U} = U \Psi  $. In particular, if $\| \Psi \| =1$, $ \pUPsiUPsi (t , \mbf{x})$ is    the solution.
\end{theorem}

%%%%%%%%%%%%%%%%%%%%%%%%%%%%%%%%%%%%%%%%%%%%%%%%%%%%%%%%%%%%%%%%%%%%%%%%%%
%%%%%%%%%%%%%%%%%%%%%%%%%%%%%%%%%%%%%%%%%%%%%%%%%%%%%%%%%%%%%%%%%%%%%%%%%%
%%%%%%%%%%%%%%%%%%%%%%%%%%%%%%%%%%%%%%%%%%%%%%%%%%%%%%%%%%%%%%%%%%%%%%%%%%
%%%%%%%%%%%%%%%%%%%%%%%%%%%%%%%%%%%%%%%%%%%%%%%%%%%%%%%%%%%%%%%%%%%%%%%%%%
\section{Proof of Theorem \ref{MainTheorem}} 

\subsection{Operator kernel of annihilation operator} \label{OperatorKernel}
The operator kernel of annihilation operator is defined by 
\[
(a(\mathbf{k}) \Psi )^{\, (n)} ( \mbf{k}_1 , \cdots , \mbf{k}_n)
= \sqrt{n+1}  \Psi ^{\, (n+1)} (\mbf{k} ,\mbf{k}_1 , \cdots , \mbf{k}_n   ), \quad n \geq 0 . 
\]
 Let   $\ms{D}_{\textrm{a} , \,f } = \left\{  \Psi \in \Fb \left. \frac{}{} \right| 
\int_{\Rd }|f(\mbf{k})| \,\| a (\mbf{k}) \Psi \| \dk  < \infty  \right\}$, $f \in L^2 (\Rd )$. It holds that 
\begin{equation}
 \int_{\Rd} f(\mathbf{k})^{\ast} ( \Phi ,  a (\mathbf{k}) \Psi ) \dk = ( \Phi , a (f) \Psi ) ,\quad \Phi \in \Fb, \;
\Psi \in \ms{D}_{\textrm{a} , \, f } \cap \ms{D}(a(f)).   \label{Dker}
\end{equation}
Let $T= T(\mbf{k})$ be a non-negative and Borel measurable function, and $f \in  L^{2} (\Rdk )$ such that $  T^{\, -1/2} f \in  L^{2} (\Rdk )$. Then  it follows that  for all $\Phi  \in \Fb , \; \Psi \in \ms{D}(\sqzb{T}^{1/2} ) $,
\begin{equation}
\int_{\Rd} \left| \frac{}{}  f(\mathbf{k})  \, ( \Phi ,   a (\mathbf{k}) \Psi  ) \right| \dk
\leq  \| T^{\, -1/2} f \| \,  \| \Phi \| \,   \| \sqzb{T}^{1/2} \Psi \|  
 . \label{8/19.f} 
\end{equation}
It also  holds that 
\begin{equation}
\qquad \qquad \int_{\Rd} T(\mbf{k})
 \| a( \mbf{k}) \Psi \|^2 \dk  = \|   \sqzb{T}^{1/2} \Psi \|^2 , 
 \qquad  \Psi  \in \ms{D} ( \sqzb{T}^{1/2} ) . \label{8/19.e}
\end{equation}

$\;$ \\
From (\ref{8/19.e}), the next lemma immediately follows. 
\begin{lemma} \label{L8/09.1} \normalfont 
Assume \textbf{(A.1)}. Then, for all $\Phi \in \Fb$ and $ \Psi \in \ms{D}(H )$,
\[
\left( \int_{\Rd}  \omega (\mbf{k})| \FPhiPsi (t, \mbf{k})|^2 \dk \right)^{1/2} \leq
\| \Phi \| \, 
\| H_{0}^{1/2} e^{\,-itH } \Psi  \| . 
\]
\end{lemma}

%%%%%%%%%%%%%%%%%%%%%%%%%%%%%%%%%%%%%%%%%%%%%%%%%%%%%%%%%%%%%%%%%%%%%%%%%%%%
%%%%%%%%%%%%%%%%%%%%%%%%%%%%%%%%%%%%%%%%%%%%%%%%%%%%%%%%%%%%%%%%%%%%%%%%%%%%
$\;$ \\
Let
\[
 \FzeroPhiPsi (\mbf{k}) = (\Phi , a(\mbf{k}) \Psi )  .
\]
\begin{proposition} \label{P8/29.1}  \normalfont
Assume \textbf{(A.1)} and \textbf{(A.2)}. Then it holds that for all $ \Phi , \Psi \in \ms{D} (H)$, 
\[
\FUPhiUPsi (t, \mbf{k})= e^{\, -it \omega (\mbf{k})}  \FzeroPhiPsi (\mbf{k}) +
\frac{1}{\sqrt{2}} \frac{\hat{\rho} (\mbf{k})\,}{\omega (\mbf{k})^{3/2}} ,\qquad 
\textrm{a.}\textrm{e.} \; \; \mbf{k}  \in \Rd . 
\]
\end{proposition}
\textbf{(Proof)} 
 Let  $ \Phi , \Psi \in \ms{D} (H)$. We set $\FOmegaUPhiUPsi (t, \mbf{k}) = \sqrt{\omega ( \mbf{k} )} \FUPhiUPsi (t, \mbf{k})$. For all
 $h \in L^2 (\Rd ) $ such that $ \sqrt{\omega} h \in L^2 (\Rd )$, we have 
\begin{align}
\int_{\Rd} h(\mbf{k})^{\ast} \FOmegaUPhiUPsi (t, \mbf{k}) \dk 
& = \int_{\Rd} \sqrt{\omega(\mbf{k})}  h(\mbf{k})^{\ast} (U \Phi , e^{\, it H} a(\mbf{k}) e^{\, - it H} U \Psi )  \dk
\notag \\
& = ( e^{\, - it H} U \Phi ,a( \sqrt{\omega} h)  e^{\, - it H} U  \Psi ) \notag \\
& = ( U^{\ast}e^{\, - it H} U \Phi , \left( U^{\ast} a(\sqrt{\omega} h) U \right) U^{\ast} e^{\, - it H} U  \Psi ) \label{8/23.0}
\end{align}
By   (\ref{diagH}),  
\begin{equation}
U^{\ast} e^{ - it  H} U = e^{ - it U^{\ast} H U}
= e^{\frac{i t}{2} (\frac{\fI}{\omega} , \fI )} \, e^{\, - it H_{0}} . \label{8/23.e} 
\end{equation}
From  (\ref{WCR1}), 
\begin{equation}
 U^{\ast} a (\sqrt{\omega} h) U =  a(  \sqrt{\omega} h) +  \frac{1}{\sqrt{2}}(  h ,  \frac{\fI}{\sqrt{\omega}}   )  . \label{8/23.f} 
\end{equation}
By  (\ref{8/23.e}), (\ref{8/23.f}) and $(\Phi , \Psi ) =1$, we have
\begin{align}
  ( U^{\ast}e^{\, - it H} U \Phi , \left( U^{\ast} a(\sqrt{\omega} h) U \right) U^{\ast} e^{\, - it H} U  \Psi )
& = \left(  \Phi, e^{\,it H_{0}}   a( \sqrt{\omega} h) e^{\, - it H_{0}} \Psi \right)  +  \frac{1}{\sqrt{2}}( h , \frac{\fI}{\sqrt{\omega}}   )  \notag \\
& = (\Phi , a ( \sqrt{\omega} e^{\, it \omega } h  ) \Psi ) +  \frac{1}{\sqrt{2}}( h , \frac{\fI}{\sqrt{\omega}}) . \notag  
\end{align}
Here we used (\ref{8/23.a}) in the last line. Then we have
\[
\int_{\Rd} h(\mbf{k})^{\ast} \FOmegaUPhiUPsi (t, \mbf{k}) \dk =
\int_{\Rd} h(\mbf{k})^{\ast}  
\left(  \sqrt{\omega (\mbf{k})} e^{\, -it \omega (\mbf{k}) } ( \Phi , a(\mbf{k}) \Psi ) +  \frac{1}{\sqrt{2}} \frac{\fI (\mbf{k})}{\sqrt{\omega (\mbf{k})}}   \right)
 \dk .
\]
Since  the set which consists of all vectors $h \in L^2 (\Rd ) $ such that $ \sqrt{\omega} h \in L^2 (\Rd )$ is dense in $ L^2 (\Rd )$ and $ \int_{\Rd} | \FOmegaUPhiUPsi (t, \mbf{k})  |^2 \dk  = \int_{\Rd}  \omega (\mbf{k})| \FPhiPsi (t, \mbf{k})|^2 \dk< \infty $ from Lemma \ref{L8/09.1},  we have
\[
\qquad \qquad 
\FOmegaUPhiUPsi (t, \mbf{k})= \sqrt{\omega (\mbf{k})} e^{\, -it \omega (\mbf{k})} (\Phi , a(\mbf{k}) \Psi) +	
\frac{1}{\sqrt{2}} \frac{\hat{\rho} (\mbf{k})}{\omega (\mbf{k})}  ,\quad \quad 
\textrm{a.}\textrm{e.} \; \; \mbf{k}  \in \Rd .
\]
By dividing both sides of the  above equation by  $\sqrt{\omega(\mbf{k})}$, the proof is obtained. $\blacksquare $ \\

%%%%%%%%%%%%%%%%%%%%%%%%%%%%%%%%%%%%%%%%%%%%%%%%%%%%%%%%%%%%%%%%%%%%%%%%%%%%%
%%%%%%%%%%%%%%%%%%%%%%%%%%%%%%%%%%%%%%%%%%%%%%%%%%%%%%%%%%%%%%%%%%%%%%%%%%%%%
%%%%%%%%%%%%%%%%%%%%%%%%%%%%%%%%%%%%%%%%%%%%%%%%%%%%%%%%%%%%%%%%%%%%%%%%%%%%%
%%%%%%%%%%%%%%%%%%%%%%%%%%%%%%%%%%%%%%%%%%%%%%%%%%%%%%%%%%%%%%%%%%%%%%%%%%%%%
\begin{lemma} \label{L7/31.2}\normalfont 
Let $T= T(\mbf{k})$ be a non-negative and Borel measurable function.
 Then, for all $p \in \mbb{N}$,
\[
\qquad \qquad  \qquad 
\| \sqzb{T^{\, p} }^{1/2} \Psi \|  \leq \| \sqzb{T}^{p/2} \Psi \|^2 ,
\qquad  \Psi  \in \ms{D} (\sqzb{T}^{p/2}) . 
\] 
\end{lemma}
\textbf{(Proof)}   Let  $ \Psi= \{ \Psin \}_{n=0}^{\infty} \in \ms{D} (\sqzb{T}^{p/2})$. We see that
{\small
\[
\int_{\mathbb{R}^{d n}}   \left( \sum_{j=1}^n T (\mbf{k}_j)^p \right)   \left|  \Psi^{\, (n )} (\mathbf{k}_1 , \cdots , \mathbf{k}_n ) \right|^2 \;
  \, \dk_{1} \cdots \dk_n 
 \leq \int_{\mathbb{R}^{d n}} \left( \sum_{j=1}^n T (\mbf{k}_j) \right)^p   \left|  \Psi^{\, (n )} (\mathbf{k}_1 , \cdots , \mathbf{k}_n ) \right|^2 \;
  \, \dk_{1} \cdots \dk_n .
\]}
From this inequality,  it is directly proven  that 
$ \| \sqzb{T^{\, p} }^{1/2} \Psi \| \leq \| \sqzb{T}^{p/2} \Psi \|  $. 
 $\blacksquare $ \\

$\;$ \\
From (\ref{8/19.e}) and   Lemma \ref{L7/31.2}, next corollary follows. 
 
\begin{corollary} \label{7/31.3}  
\normalfont 
Let $p \in \mbb{N}$. Then,
\[
\qquad \qquad  \qquad 
\int_{\Rd} \omega (\mbf{k})^p
 \| a( \mbf{k}) \Psi \|^2 \dk 
  \leq \| H_0^{p/2} \Psi \|^2  , \qquad \Psi \in \ms{D} (H_0^{p/2}) .
\]
\end{corollary}

\begin{proposition} \label{P7/31.3} \normalfont
Assume \textbf{(A.1)} and  $d \geq 3$.    Then  for all   $l=0, 1 ,2$ and $ j=1, \cdots , d $,
\[
\int_{\Rd} \frac{|k_{j} |^l}{ \sqrt{\omega (\mbf{k})}}
| \FzeroPhiPsi (\mbf{k}) | \dk \,  \leq \; c_{d}
\| \Phi  \| \left( 
 \| H_0^{\frac{d }{2} +2}  \Psi \| + \|   \Psi \|  \right) , \quad  \Psi \in \ms{D} (H_{0}^{\frac{\,  d  }{2} + 2}) ,
 \]
\end{proposition}
where $c_{d} =  \left\| \frac{1}{\omega} \right\|_{L^2 (\Bd )} + \left\| \frac{1}{\omega^{\frac{d+1}{2}}} \right\|_{L^2 ( \Bd^{\, c} )}$. 
$\;$ \\
\textbf{(Proof)} 
Let  $ \Psi \in \ms{D} (H_{0}^{\frac{\,  d  }{2} + 2} ) $.
We see that  
\[
 \int_{\Rd} \frac{|k_{j} |^l}{ \sqrt{\omega (\mbf{k})}}
 | \FzeroPhiPsi (\mbf{k}) |  \dk 
\leq \|  \Phi \| \, \int_{\Rd} \frac{|k_{j} |^l}{ \sqrt{\omega (\mbf{k})}}
\|a (\mbf{k})  \Psi \| \dk 
\]
Note that      $ \int_{\Bd} \frac{1}{\omega(\mbf{k})^2} \dk < \infty  $, for   all $d \geq 3$. Then, 
\begin{align}
\int_{\Bd} \frac{|k_{j} |^l}{ \sqrt{\omega (\mbf{k})}}
\|a (\mbf{k})  \Psi \| \dk &  \leq 
\int_{\Bd} \frac{1}{ \sqrt{\omega (\mbf{k})}}
\|a (\mbf{k})  \Psi \| \dk  \notag \\
& \leq  
 \left\| \frac{1}{\omega} \right\|_{L^2 (\Bd )} \left(
\int_{\Bd} \omega (\mbf{k}) 
\|a (\mbf{k})  \Psi \|^2 \dk  \right)^{1/2} \notag  \\
& \leq  \left\| \frac{1}{\omega} \right\|_{L^2 (\Bd )}
\| H_0^{1/2}  \Psi \| .  \label{8/09.a} 
\end{align}
We also note that $\int_{\Rd \backslash \Bd} \frac{1}{\omega (\mbf{k})^{d+1 } } \dk < \infty $ for   all $d \in \mbb{N}$.  Let $p_{d,l} = d +2l$, $l=0, 1 ,2$. Then,
\begin{align}
\int_{\Bd^{\, c} } \frac{|k_{j} |^l}{ \sqrt{\omega (\mbf{k})}}
\|a (\mbf{k}) \Psi \| \dk &  \leq 
\int_{ \Bd^{\, c} } \frac{ \omega (\mbf{k})^l}{ \sqrt{\omega (\mbf{k})}}
\|a (\mbf{k})  \Psi \| \dk \notag  \\
& \leq  
\left\| \frac{1}{\omega^{\frac{p_{d ,l} +1}{2}-l}} \right\|_{L^2 (\Bd^{\, c} )}
 \left(
\int_{\Rd \backslash \Bd } \omega (\mbf{k})^{p_{d, l} } 
\|a (\mbf{k}) \Psi \|^2 \dk  \right)^{1/2} \notag \\
& \leq    \left\| \frac{1}{\omega^{\frac{d+1}{2}}} \right\|_{L^2 ( \Bd^{\, c} )}
\| H_0^{\frac{p_{d,l}}{2}}  \Psi \|.  \label{8/09.b} 
\end{align}
From (\ref{8/09.a}) and (\ref{8/09.b}), we have 
\begin{equation}
\int_{\Rd} \frac{|k^{j} |^l}{ \sqrt{\omega (\mbf{k})}}
\|a (\mbf{k})  \Psi \| \dk  
\leq \left\| \frac{1}{\omega} \right\|_{L^2 (\Bd )}
\| H_0^{1/2}  \Psi \|
+ 
\left\| \frac{1}{\omega^{\frac{d+1}{2}}} \right\|_{L^2 ( \Bd^{\, c} )}
\| H_0^{\frac{p_{d,l}}{2}}  \Psi \| . \label{8/09.c}
\end{equation}
By  spectral decomposition theorem,     $ \| H_{0}^{1/2} \Psi \| \leq  \| H_0^{\frac{d+4}{2}}  \Psi \| + \| \Psi \| $ and $\| H_0^{\frac{p_{d,l}}{2}}  \Psi \| \leq \| H_0^{\frac{d+4}{2}}  \Psi \| + \| \Psi\|$, $l=0,1,2$. Hence  we obtain the proof. $\blacksquare $ \\

%%%%%%%%%%%%%%%%%%%%%%%%%%%%%%%%%%%%%%%%%%%%%%%%%%%%%%%%%%%%%%%%%%%%%%%%%%%%%%%%%%%%%%%%%%%%%%%%%%%%%%%%%%%%%%%%%%%%%%%%%%%%%%%%%%%%%%%%%%%%%%%%%%%%%%%%%%%%%%%%%%%%%%%%%%%%%%%%%%%%%%%%%%%%%%

$\;$ \\
{\large\textbf{(Proof of Theorem \ref{MainTheorem})}} \\ 
Let $ \Phi , \Psi  \in  \ms{D} (H_{0}^{\frac{\,  d  }{2} + 2})  $. 
From Proposition \ref{P8/29.1}, Proposition \ref{P7/31.3} for $l=0$ and \textbf{(A.2)},
we have  
\begin{align}
\pUPhiUPsi (t, \mbf{x})
 =&  \frac{1}{\sqrt{2 \pi}^{\, d}} \int_{\Rd} 
\frac{1}{\sqrt{2 \omega (\mbf{k})}} 
\left( e^{ \, -i t  \omega ( \mbf{k})  +  i \,\mbf{k} \cdot \mbf{x} } 
\FzeroPhiPsi (\mbf{k}) + e^{ \, i t  \omega ( \mbf{k}) - i\,   \mbf{k} \cdot \mbf{x} } 
 \FzeroPsiPhi (\mbf{k})^{\ast} 
 \right) \dk   \notag \\
 & \qquad \qquad  +  \frac{1}{\sqrt{2 \pi}^{\, d}} \int_{\Rd} \frac{\hat{\rho} ( \mbf{k}) }{\omega(\mbf{k})^2} 
 e^{i \mbf{k} \cdot \mbf{x}}  \dk .  \notag
\end{align}
By  Proposition \ref{P7/31.3} for $l=1,2$, we have  
\begin{equation}
\partial_t^2 \pUPhiUPsi (t, \mbf{x}) =-\frac{1}{\sqrt{2 \pi}^{\,d}} 
\int_{\Rd} \frac{\omega(\mbf{k})^2}{\sqrt{2 \omega (\mbf{k})}} 
\left( e^{ \, -i t \omega  ( \mbf{k})  + i \,   \mbf{k} \cdot \mbf{x} } 
 \FzeroPhiPsi  (\mbf{k})+ e^{ \, i t \omega  (\mbf{k} )  -  i\,  \mbf{k} \cdot \mbf{x} } 
  \FzeroPsiPhi(\mbf{k})^{\ast}  \label{8/23.i}
 \right) \dk  , \notag 
\end{equation}
and
\begin{align*}
  \triangle  \,  \pUPhiUPsi (t, \mbf{x})  
&=  \frac{-1}{\sqrt{2 \pi}^{\,d}} \sum_{j=1}^d \left\{ \int_{\Rd}  
\frac{{k}_{j}^2 }{\sqrt{2 \omega (\mbf{k})}} 
\left( e^{ \, -i t \omega ( \mbf{k}) +i \,  \mbf{k} \cdot \mbf{x} } 
\FzeroPhiPsi (\mbf{k}) + e^{ \, i t \omega ( \mbf{k})  - i \mbf{k} \cdot \mbf{x} } 
  \FzeroPsiPhi (\mbf{k})^{\ast}  \right)  \right. \dk    \\
 & \qquad \qquad \qquad  \qquad +   \left. \frac{1}{\sqrt{2 \pi}^{\,d}} \int_{\Rd} \frac{k_{j}^2 }{\omega (\mbf{k})^2} \hat{\rho} ( \mbf{k})   e^{\, i \, \mbf{k} \cdot \mbf{x}}  \dk \right\}    \\
&=  \frac{-1}{\sqrt{2 \pi}^{\,d}}   \int_{\Rd}  
\frac{\omega (\mbf{k})^2}{\sqrt{2 \omega (\mbf{k})}} 
\left( e^{ \, -i t \omega ( \mbf{k}) +i \,  \mbf{k} \cdot \mbf{x} } 
\FzeroPhiPsi (\mbf{k}) + e^{ \, i t \omega ( \mbf{k})  - i\mbf{k} \cdot \mbf{x} } 
  \FzeroPsiPhi (\mbf{k})^{\ast}  \right)  \dk    \\
 & \qquad \qquad \qquad  \qquad -   \frac{1}{\sqrt{2 \pi}^d} \int_{\Rd}  \hat{\rho} ( \mbf{k})   e^{\, i \, \mbf{k} \cdot \mbf{x}}  \dk    \\
 &  = \partial_t^2 \pUPhiUPsi (t, \mbf{x}) - \rho (\mbf{x}) . 
 \end{align*}
Then we have $  ( \partial_t^2  - \triangle ) \pUPhiUPsi (t, \mbf{x}) = \rho (\mbf{x}) $.  $\blacksquare $

$\;$ \\
\textbf{[Concluding remark]}\\
Let us consider   massive cases $\omega_{\, m} (\mbf{k}) = \sqrt{\mbf{k}^2 + m^2}$, $m>0$. Similarly, we can  construct the solutions of $ (\partial_t^2 - \triangle +m^2 ) \phi_{\textrm{cl}} (t , \mbf{x}) = \rho (\mbf{x})$.  In these cases we do not need to suppose $d \geq 3$, since   $\left\| \frac{1}{\omega_{\, m} (\mbf{k})} \right\|_{L^2 (\Bd )} < \infty $,  $d \in \mbb{N}$, which   is  correspond to   $\left\| \frac{1}{\omega (\mbf{k})} \right\|_{L^2 (\Bd )}  < \infty $, $ d \geq 3$,  in Proposition \ref{P7/31.3}.\\

$\;$ \\
{\large \textbf{Acknowledgments}}  
This work is supported by JSPS grant $16$K$17607$.

\end{document}